# $L_4Fe_2As_2Te_{1-x}O_{4-y}F_y$ (L = Pr, Sm, Gd): a layered oxypnictide superconductor with $T_c$ up to 45 K

S. Katrych[1], A. Pisoni[1], S. Bosma[2], S. Weyeneth[2], N. D. Zhigadlo[3], R. Gaal[1], J. Karpinski[1,3], L. Forró[1]

[1]*Institute of Condensed Matter Physics, EPFL, CH-1015 Lausanne, Switzerland*
[2]*Physik-Institut der Universität Zürich, Winterthurerstrasse 190, CH-8057 Zürich, Switzerland*
[3]*Laboratory for Solid State Physics, ETH Zurich, CH-8093 Zurich, Switzerland*

**Abstract**

The synthesis, structural and physical properties of iron lanthanide oxypnictide superconductors, $L_4Fe_2As_2Te_{1-x}O_4$ (L = Pr, Sm, Gd), with transition temperature at ~ 25 K are reported. Single crystals have been grown at high pressure using cubic anvil technique. The crystal structure consists of layers of $L_2O_2$ tetrahedra separated by alternating layers of chains of Te and of $Fe_2As_2$ tetrahedra: $-L_2O_2-Te-L_2O_2-Fe_2As_2-L_2O_2-Te-L_2O_2-$ (space group: *I*4/*mmm*, $a$ ~ 4.0, $c$ ~ 29.6 Å). Substitution of oxygen by fluorine increases the critical temperature, e.g. in $Gd_4Fe_2As_2Te_{1-x}O_{4-y}F_y$ up to 45 K. Magnetic torque measurements reveal an anisotropy of the penetration depths of ~31.

## I. INTRODUCTION

The discovery of high temperature superconductivity in $LaFeAsO_{1-x}F_x$[1] has spiralled up an investigation of a variety of compounds consisting of $M_2As_2$ (M is a transition metal) layers alternating with $L_2O_2$ (L is a lanthanide metal) layers or metal layers (eg. Ba, Li or other). In the view of the fact that all of these compounds were reported already before discovery of superconductivity in La1111[2,3] and their synthesis procedure, as well as structure details were well investigated, great progress in finding of several new iron based superconductors was achieved by doping of known layered pnictide and oxypnictide in a less than one year. Superconductivity was also found in FeSe or "11" ($T_c$ = 8 K)[4,5], $Ba_{1-x}K_xFe_2As_2$ or "122" ($T_c$ = 38 K)[6,7], LiFeAs or "111" ($T_c$ = 18 K)[8,9], $La_3Ni_4P_4O_2$ or "3442" ($T_c$ = 2.2 K)[10,11].

In this work we present the systematic study of superconducting compounds $L_4Fe_2As_2Te_{1-x}O_4$ (L = Pr, Sm, Gd) with a hitherto unknown structure. Our recent work[12], where crystal growth, structure and physical properties of an oxypnictide superconductor $Pr_4Fe_2As_2Te_{1-x}O_4$ with a critical temperature ($T_c$) close to 25 K were reported, triggered this study with the intention of increasing $T_c$ of this compound.

The structure of $Pr_4Fe_2As_2Te_{1-x}O_4$ (also called 42214) reveals the same structural blocks as $PrFeAsO$ (1111) oxypnictide superconductor[3]: fluorite-type $Pr_2O_2$ layers alternating with anti-fluorite-type $Fe_2As_2$ layers with a difference in a presence of intercalated Te atoms (Fig. 1a-c). Both structures are tetragonal with the space group $I4/mmm$ for the former and $P4/nmm$ for the latter.

For the 1111 phase two parameters are known how to increase $T_c$ by modifying the structure or composition. One of them is pnictogen height $h_P$ (half of the thickness of the $Fe_2As_2$ or $S_2$, Fig. 1a-c ), which as indicator for highest possible $T_c$ should be close to 1.38 Å[13, 14]. The LFeAsO for L = Sm and Gd have $h_P$ value close to the "optimal" one.

The other factor is doping. The pure stoichiometric, undoped 1111 phase does not reveal superconductivity[1]. The most effective doping that results in $T_c$ above 50 K is at the oxygen sites either by fluorine, hydrogen or by oxygen vacancies[1, 15, 16]. Substitution of Sm by Th also leads to $T_c$ above 50 K[17, 18].

According to our recent investigations[12, 19] the lattice constant $a$, $b$, $Fe_2As_2$ layer thickness $S_2$ and $Fe_2As_2$-$L_2O_2$ interlayer distance $S_3$ are almost equal for both PrFeAsO and for $Pr_4Fe_2As_2Te_{1-x}O_4$ (Fig. 1a-c). In analogy to 1111 one can expect that "optimal" $h_P$ and proper doping should lead to high $T_c$ in 42214 compounds. $L_4Fe_2As_2Te_{1-x}O_4$ (L = Sm and Gd) doped with fluorine, as one of the options, was assumed as the ideal candidate for increasing $T_c$ in 42214 because $h_P$ of LFeAsO (L=Sm, Gd) corresponds to the maximal $T_c$. To prove this suggestion a detailed structural study of undoped $L_4Fe_2As_2Te_{1-x}O_4$ (L = Pr, Sm, Gd) was preformed as well as the influence of fluorine substitution for oxygen on $T_c$ in $L_4Fe_2As_2Te_{1-x}O_4$ (L = Sm, Gd) was studied.

Here we report high pressure crystal growth, structure and doping of $L_4Fe_2As_2Te_{1-x}O_{4-y}F_y$ (L = Pr, Sm, Gd) single crystals. The critical temperature of Gd42214 single crystals increased after F doping from 25 K up to 45 K.

## II. EXPERIMENTAL DETAILS

The crystals of the $Pr_4Fe_2As_2Te_{1-x}O_4$, $Sm_4Fe_2As_2Te_{1-x}O_{4-y}F_y$ and $Gd_4Fe_2As_2Te_{1-x}O_{4-y}F_y$ were grown at high pressure from precursors (pre-sintered mixture of 0.9PrAs + 0.1PrTe + FeO, 0.9SmAs + 0.1TeO_2 + 0.5FeO + 0.15FeF_2 + 0.35Fe + 0.1Sm and Fe + 0.9GdAs + 0.1GdTe + 0.8FeO + 0.1FeF_2, respectively) in NaCl/KCl flux. The procedure is described by Katrych *et al.*[12] The same procedure was used for the growth of $L_4Fe_2As_2Te_{1-x}O_4$ (L = Sm, Gd) crystals from stoichiometric mixture of high purity (≥ 99.95%) L, LAs, FeO and $TeO_2$.

The single crystals were studied on a 3 circle x-ray diffractometer equipped with a CCD detector (Bruker AXS Inc.). Data reduction and multi-scan absorption correction were performed using the APEX2[20] and SAINT[21] software. The crystal structure was solved by a direct method and full data set refined on $F^2$, employing the programs SHELXS-97[22] and SHELXL-97[23].

The magnetization as a function of temperature has been measured for individual single crystals of $L_4Fe_2As_2Te_{1-x}O_4$ (L = Pr, Gd) and $L_4Fe_2As_2Te_{1-x}O_{4-y}F_y$ (L = Sm, Gd) doped with fluorine. These experiments have been done with an MPMS with enhanced sensitivity (MPMS-XL).

Because individual $Sm_4Fe_2As_2Te_{1-x}O_4$ single crystals were too small, the magnetic susceptibility ($M/H$) measurement was performed on a powdered sample using a magnetic property measurement system (MPMS: Quantum Design) under zero-field-cooling (ZFC) and field-cooling (FC) conditions.

The electrical resistivity ($\rho$) of a $Sm_4Fe_2As_2Te_{1-x}O_{4-y}F_y$ single crystal was measured as a function of temperature using a standard four-probe method. Gold wires of 12 $\mu$m diameter were glued with silver paint on a 150 $\mu$m long sample in a van der Pauw geometry. The contact resistance was less than 200 $\Omega$. A sensing current of 100 $\mu$A was generated using a Keithley 220 programmable current source and the resulting signal was measured by a Keithley 2182 nanovoltmeter. The sample was cooled in liquid helium bath cryostat from room temperature at a rate of 0.5 K/min.

The magnetic anisotropy measurements were carried out with a homemade magnetic torque sensor [24]. A tiny $Sm_4Fe_2As_2Te_{1-x}O_{4-y}F_y$ single crystal (~ 100 x 100 x 10 $\mu m^3$) was fixed on a platform hanging on piezoresistive legs. When a static magnetic field was applied to this anisotropic superconductor, a torque proportional to the vector product of magnetization and field appears and bends the piezoresistive legs. The resulting change in resistance can be read out electronically, yielding a signal proportional to the torque[24].

## III RESULTS AND DISCUSSION

### A. Structure analysis

The crystals of $Pr_4Fe_2As_2Te_{1-x}O_4$ with plate-like shape reveal nice quality with a mosaic spread of ~1.1 ° (Fig. 2).

Structure of 42214 (Fig. 1(a)) was determined and refined on the base of 670 independent reflections. Since the displacement parameter of Te was larger than that of others atoms, the occupation parameter for it was set as a free variable. Finally Te site reveals about 10 % of vacancies[12]. The same Te deficiency was found in $L_4Fe_2As_2Te_{1-x}O_4$ (L = Sm, Gd) which was refined using the $Pr_4Fe_2As_2Te_{1-x}O_4$ structural model. Table 1 and table 2 provide the results of structure determination and refinement.

Despite the entirely different stoichiometric composition or formula type, the determined structure has some resemblance to that of $La_3Cu_4P_4O_2$[10] or another representative of its structure type, superconducting $La_3Ni_4P_4O_2$ (3442)[11]. The number of atoms in the unit cells as well as the space group are the same (Pearson symbol: $tI26$, space-group: $I4/mmm$). Like in the 3442 the structure of 42214 consists of stacked fluorite-type $L_2O_2$ and anti-fluorite-type $M_2Pn_2$ layers. However, the sequence and numbers of 2 types of layers are dissimilar for the two structures, furthermore the atom in the centre of unit cell has positive charge $L^{\delta+}$ for 3442 and negative $Te^{\delta-}$ for 42214. The two structures are isoconfigurational and have interchanged corresponding structure motifs: $(L_2O_2)_2(M_2Pn_2)Te$ vs. $(L_2O_2)(M_2Pn_2)_2L$ (Fig. 1a-e), where L = Pr, Sm, Gd and M = Fe; Pn = As for the former and L = La, Ce, Nd; M = Cu, Ni; Pn = P for the latter. Regarding the definition of Lima-de-Faria et al [25], the 42214 structure could be considered as an *anti* - $La_3Cu_4P_4O_2$ – crystal structure.

The 3442 phase is considered as an ordered mixture of LMPnO and $LM_2Pn_2$[11] (Fig.2c-d). The 42214 could be described as stacking of the 1111 with $L_2O_2Te$[26] (Fig. 1a-b-c).

Across the lanthanide series with increasing the number of atom in periodic table and decreasing of its covalent radius lattice constants, thickness of $Pr_2O_2$ layers $S_1$ decreases while thickness of $Fe_2As_2$ layer, $S_2$ (2 x $h_P$), increases (Fig. 3). The corresponding $S_2$ and $S_3$ are very close for both 1111 and 42214 structures[19] (Tab. 1, Fig. 1a-c) while the $L_2O_2$ ($S_1$) layer is by 0.1 Å thicker in 1111. The $h_P$ parameter increases monotonically from 1.332 to 1.366 Å and is getting, as it was expected, close to "optimal" value for $Sm_4Fe_2As_2Te_{1-x}O$ and $Gd_4Fe_2As_2Te_{1-x}O$ (1.352(3) and 1.366(3) Å, respectively).

The fluorine-doped crystals of $Sm_4Fe_2As_2Te_{1-x}O_{4-y}F_y$ and $Gd_4Fe_2As_2Te_{1-x}O_{4-y}F_y$ were also studied by single crystal x-ray diffraction. The results are shown in tables 3 and 4. In both cases the lattice parameter $c$ is smaller than for samples without fluorine. For $Sm_4Fe_2As_2Te_{1-x}O_{4-y}F_y$ $c$ is even much smaller that could be caused by a high concentration of tellurium deficiency (about 28 *at. %*). The lattice constants $a$ and $b$ for $Gd_4Fe_2As_2Te_{1-x}O_{4-y}F_y$ remain almost the same as for sample without fluorine. For $Sm_4Fe_2As_2Te_{1-x}O_{4-y}F_y$ $a$ and $b$ are slightly smaller in comparison to $Sm_4Fe_2As_2Te_{1-x}O_4$. It is rather difficult to find any correlation or exact influence of fluorine doping with the lattice geometry because of different concentration of vacancies in Te site for each sample as well as of difficulties with estimation of the real fluorine content.

**B. Magnetic properties**

*a) Samples with Te deficiency*

The temperature dependence of the magnetic moment for the $Sm_4Fe_2As_2Te_{1-x}O_4$ powdered sample (m = 0.02 g) and for two small individual single crystals of $L_4Fe_2As_2Te_{1-x}O_4$ (L = Pr, Gd) is shown in figures 4 and 5, respectively. The $M(T)$ measurements for single crystals have been performed in various magnetic fields parallel to the *c*-axis, for the zero-field-cooling (ZFC) and field-cooling (FC) states. The observed signal reveals bulk superconductivity[12]. Magnetic measurements performed on $Sm_4Fe_2As_2Te_{1-x}O_4$ powdered sample at applied field of 0.5 mT showed the onset of superconducting transition at ~ 25 K with relatively small superconducting shielding fraction (9.4 %, see the ZFC curve in Fig. 4, the density of the compound was calculated to be 7.48 g/cm$^3$). The powdered sample was not single phase.

The onset for $Sm_4Fe_2As_2Te_{1-x}O_4$ sample as well as for both single crystals of $L_4Fe_2As_2Te_{1-x}O_4$ (L = Pr, Gd) is estimated to be nearly the same, $T_c$ ~ 25 K (Fig. 4-5). The omnipresent Te deficiency could be assumed as a doping causing superconductivity. The application of small magnetic fields above of 1 mT already lead to a drastic broadening of the transition width, indicative for a very small lower critical field, $H_{c1}$.

*b) Samples with Te deficiency and doped with fluorine*

In Fig. 6 the result of fluorine doping can be seen. Fig. 6(a) shows $M(T)$ for the zero-field-cooling (ZFC) and field-cooling (FC) states of single crystal of $Sm_4Fe_2As_2Te_{0.72(1)}O_{4-y}F_y$ After substituting nominally 30% of oxygen by fluorine $T_c$ increased from 25 K of to 40 K. Fig. 6(b) shows $M(T)$ for a $Gd_4Fe_2As_2Te_{0.92(1)}O_{4-y}F_y$ single crystal. Here, substitution nominally 20% of oxygen by fluorine increased $T_c$ up to 45 K. It is not clear if maximum $T_c$ has been obtained. Doping dependence on $T_c$ should be investigated in broader range of compositions.

**C. Transport measurement**

The onset superconducting critical temperature is 36 K for $Sm_4Fe_2As_2Te_{1-x}O_{4-y}F_y$ (Fig. 7). The $\rho$-$T$ curve suggests that the sample is not optimally doped and that some inhomogeneity is present in the structure. This might explain the small upturn in $\rho$ just above the superconducting transition.

**D. Magnetic torque studies**

Figure 8 shows the torque as a function of the angle $\theta$ between the applied magnetic field $H$ and the crystallographic *c*-axis. When the field direction is swept clockwise (CW) and counterclockwise (CCW), a hysteresis between the CW and CCW torque curves appears. This is due to the pinning of the vortices on crystal defects [27]. The data shown here is the average of the CW and CCW curves.

The angular torque data are analyzed with the Kogan model [28], which allows to extract the anisotropy $\gamma$ of the material. The anisotropy parameter $\gamma$ is defined as $\lambda_c/\lambda_{ab}$, where $\lambda_i$ is the magnetic field penetration depth for a field along the crystallographic $i$-axis. The fit is done assuming a Werthamer-Helfand-Hohenberg [29] dependence of the upper critical field, with a slope at $T_c$ given in Ref. [12].

The fit results in a rather high anisotropy of 31, measured at 25 K. This anisotropy is one of the highest among the iron-based superconductor family, consistent with the large distance between the superconducting $Fe_2As_2$ layers in this material.

## IV CONCLUSION

Single crystals of layered superconductors $L_4Fe_2As_2Te_{1-x}O_4$ (L = Pr, Sm, Gd; x ~ 0.1) were grown using high-pressure cubic anvil technique. The structure reveals the stacking of anti-fluorite-type $M_2Pn_2$ and fluorite-type $L_2O_2$ layers alternating with the chain of tellurium atoms and could be considered as an *anti*-$La_3Cu_4P_4O_2$ – crystal structure. The critical temperature $T_c$ is about 25 K and does not show any dependence on the kind of lanthanide atom.

Doping with fluorine increases $T_c$ of $Sm_4Fe_2As_2Te_{1-x}O_{4-y}F_y$ up to 40 K and of $Gd_4Fe_2As_2Te_{1-x}O_{4-y}F_y$ up to 45 K. For determination of maximum $T_c$ of 42214 more detailed studies of $T_c$ dependence on doping is necessary.

This work was supported by the Swiss National Science Foundation (Project № 140760) and by the European Community FP7 Super-Iron Project.

* sergiy.katrych@epfl.ch

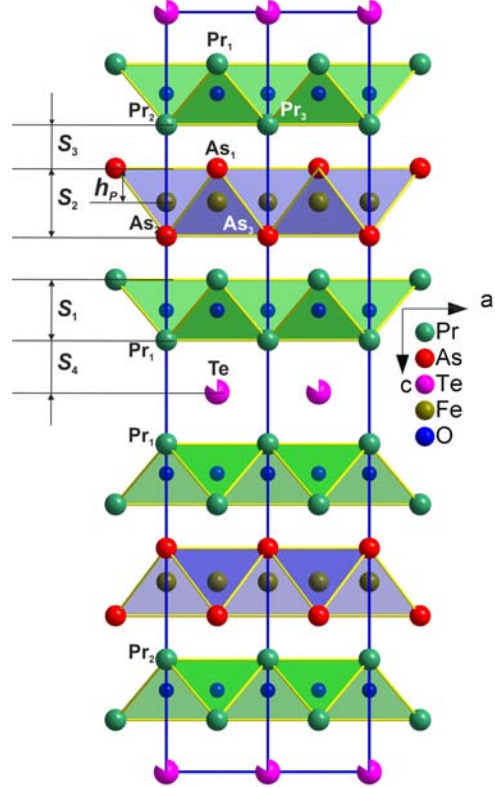 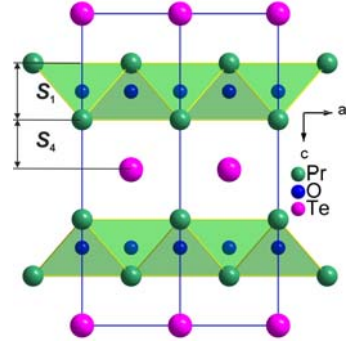 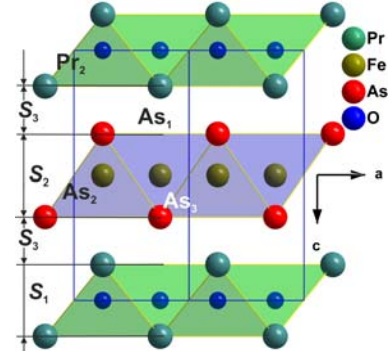 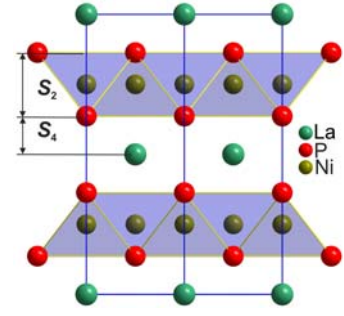 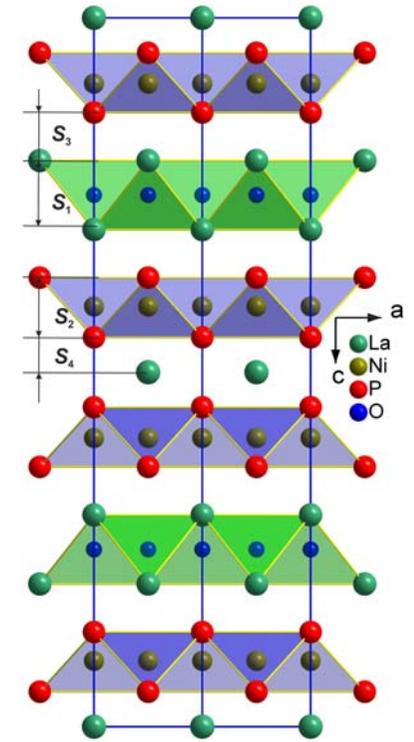

$L_4M_2Pn_2TeO_4$, L=Pr, Sm, Gd; M=Fe; Pn=As

$L_2O_2Te$; L=La-Nd, Sm-Ho[26]

LMPnO; L=La-Nd, Sm, Gd-Ho[16, 19, 30, 31]; M=Mn-Ni; Pn=P, As, Sb[32]

$MeM_2Pn_2$; Me=Ca, Sr, Ba, K, Eu, La, M= Cr-Cu, Ru, Rh, Ir Pn=P, As[33]

$L_3M_4Pn_4O_2$ L=La, Ce, Nd, M=Ni, Cu, Pn=P[10, 11]

| | | | | | | | |
|---|---|---|---|---|---|---|---|
| $S_1$=2.377(4) Å | Space group (SG): $I4/mmm$ | $S_1$=2.339 Å | SG: $I4/mmm$ | $S_1$=2.462(2) Å | SG: $P4/nmm$ | | SG: $I4/mmm$ | $S_1$=2.550 Å | SG: $I4/mmm$ |
| $1/2S_2$=$h_P$=1.332(3) | Pearson symbol (PS): t$I$26 | | PS: t$I$10 | $1/2S_2$=$h_P$=1.3429(9) | PS: t$P$8 | $1/2S_2$=$h_P$=1.21 Å | PS: t$I$10 | $S_2$=2.200 | PS: t$I$26 |
| $S_3$=1.714(4) | Lattice constants (LC): $a$~4.01; | | LC: $a$~4.06; | $S_3$=1.716(1) | LC: $a$~3.98; | $S_4$=1.19 | LC: $a$~4.01 | $S_3$=1.775 | LC: $a$~4.01; |
| $S_4$=2.042(3) | $c$~29.86 Å | $S_4$=2.045 | $c$~12.86 Å | | $c$~8.58 Å | | $c$~9.63 Å | $S_4$=1.291 | $c$~26.18 Å |

(a) (b) (c) (d) (e)

FIG. 1. (Color online) 1 x 2 unit cells projection in the *ac* plane (along *b* direction) from left to the right: (a) $L_4M_2Pn_2TeO_4$; (b) $L_2O_2Te$; (c) LMPnO; (d) $LM_2Pn_2$; (e) $L_3M_4Pn_4O_2$.

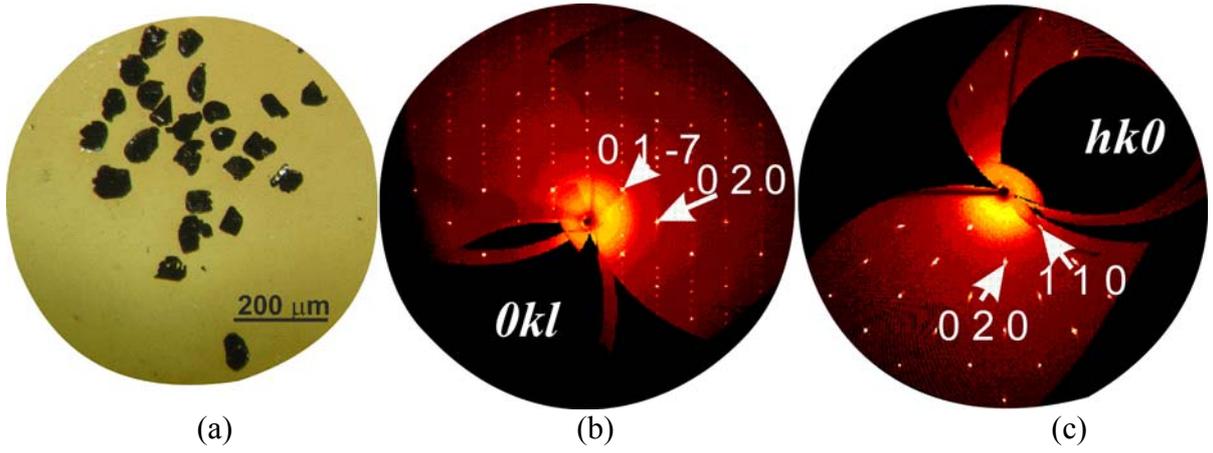

FIG. 2. (Color online) (a): single crystals of $Pr_4Fe_2As_2Te_{1-x}O_4$. (b) and (c): the reconstructed (*0kl*) and (*hk0*) reciprocal space sections for $Pr_4Fe_2As_2Te_{1-x}O_4$.

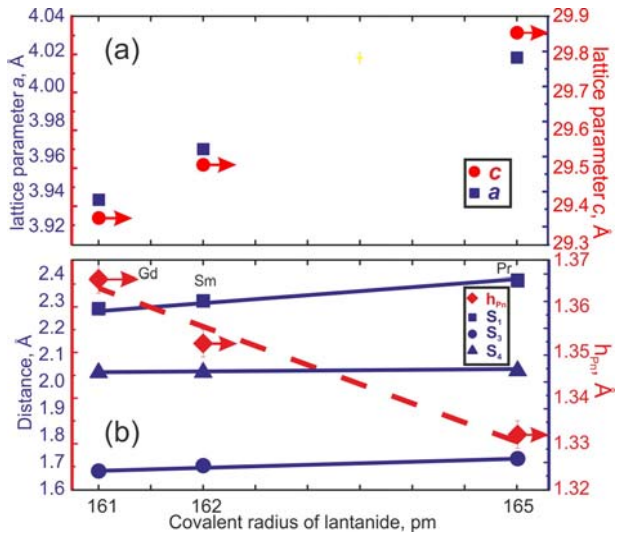

FIG. 3. (Color online) (a) and (b) - lattice parameters and layer thickness with interlayer distances as a function of covalent radius of lanthanide (arrows show that right Y axis should be used).

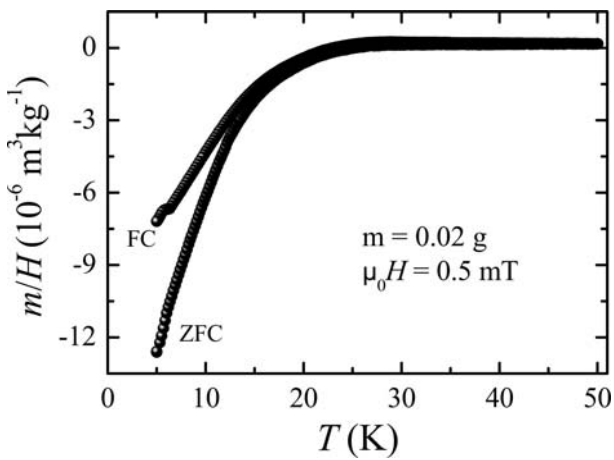

FIG. 4. Temperature dependence of the magnetic susceptibility for $Sm_4Fe_2As_2Te_{0.92(1)}O_4$ measured at 0.5 mT. FC is field cooled, ZFC is zero-field cooled.

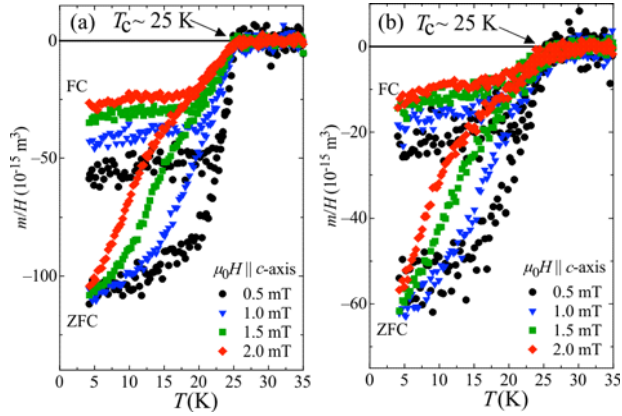

FIG. 5. (Color online) Temperature dependence of the magnetic susceptibility for:
a) $Pr_4Fe_2As_2Te_{0.88(1)}O_4$ and
b) $Gd_4Fe_2As_2Te_{0.90(1)}O_4$ single crystals.

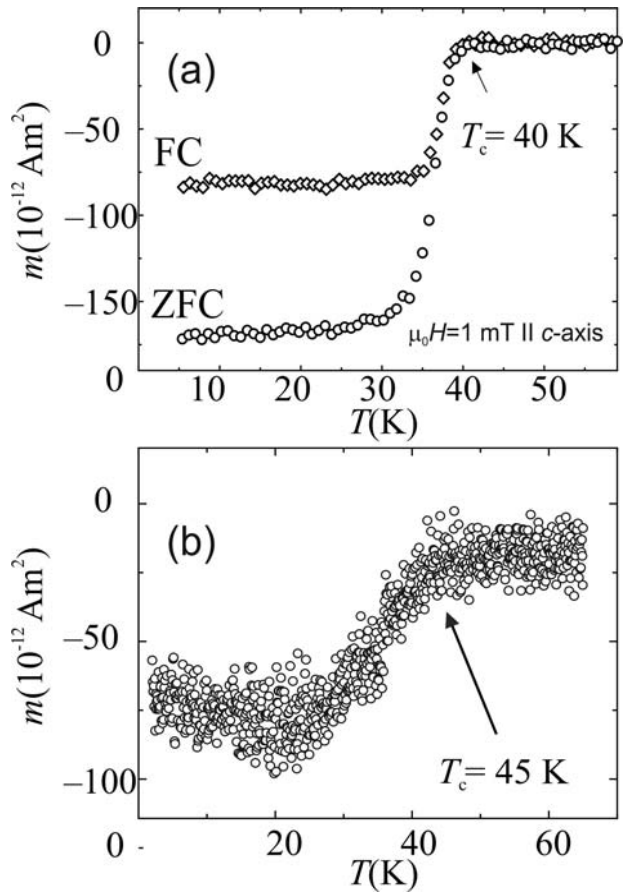

FIG. 6. Temperature dependence of the magnetization for:

a) $Sm_4Fe_2As_2Te_{0.72(1)}O_{4-y}F_y$ and

b) $Gd_4Fe_2As_2Te_{0.92(1)}O_{4-y}F_y$ single crystals.

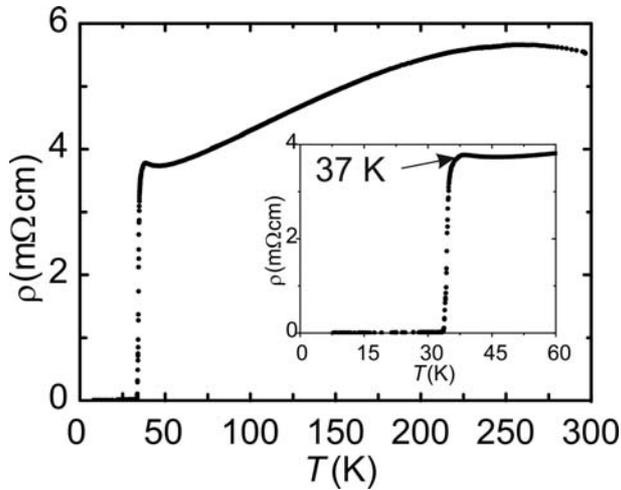

FIG. 7. Temperature dependence of the electrical resistivity for $Sm_4Fe_2As_2Te_{1-x}O_{4-y}F_y$ single crystal. Inset shows $\rho(T)$ around the superconducting transition.

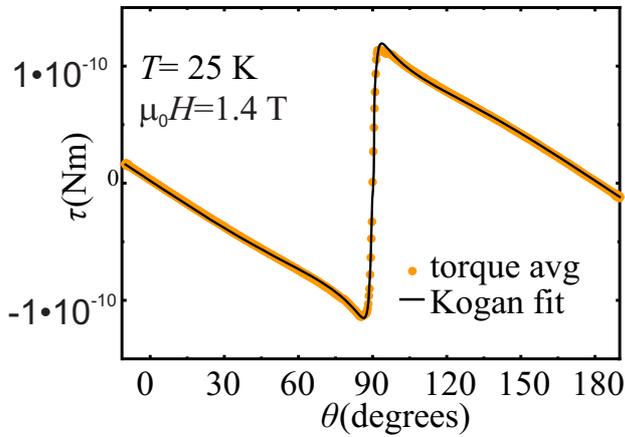

FIG. 8. (Color online) Torque as a function of the angle $\theta$ between the applied magnetic field $H$ and the crystallographic $c$-axis of a single crystal of $Sm_4Fe_2As_2Te_{0.72(1)}O_{4-y}F_y$ at $T = 25$ K and $\mu_0H = 1.4$ T. The solid line denotes a fit according to the Kogan model from which the anisotropy parameter $\gamma = 31$ was extracted.

TABLE 1. Details of the structure refinement for the $L_4Fe_2As_2Te_{1-x}O_4$ (L = Pr, Sm, Gd) crystals. The diffraction study is performed at 295(2) K using Mo $K_\alpha$ radiation with $\lambda$ = 0.71073 Å. The lattice is tetragonal, space group is *I*4/*mmm*, Z = 2. A full-matrix least-squares method was employed to optimize $F^2$.

| Empirical formula | $Pr_4Fe_2As_2Te_{0.88(1)}O_4$ | $Sm_4Fe_2As_2Te_{0.92(1)}O_4$ | $Gd_4Fe_2As_2Te_{0.90(1)}O_4$ |
|---|---|---|---|
| $T_c$, K | 25 | 25 | 25 |
| Unit cell dimensions, Å | *a*= 4.0165(2), *c*= 29.8572(16) | *a*= 3.9642(3), *c*= 29.509(2) | *a*= 3.9353(3), *c*= 29.369(3) |
| Volume, Å$^3$ | 481.66(4) | 463.74(6) | 454.82(7) |
| $L_1$-$L_2$, Å | 3.7026(6) | 3.6278(4) | 3.5928(8) |
| O-O= Fe-Fe, Å | 2.84009(14) | 2.8031(2) | 2.7827(2) |
| $L_2$-$As_1$/$L_1$-Te, Å | 3.3176(7)/3.4986(4) | 3.2718(6)/3.4638(3) | 3.2438(10)/3.4485(5) |
| $L_2$-O/$L_1$-O, Å | 2.346(2)/2.320(3) | 2.305(2)/2.2802(19) | 2.284(4)/2.261(4) |
| $As_1$-$As_2$, Å | 3.8922(12) | 3.8953(9) | 3.8992(18) |
| Fe-As, Å | 2.4091(7) | 2.3995(5) | 2.3952(10) |
| $As_1$-Fe-$As_2$, $\beta$ (°) | 107.77(2) | 108.521(17) | 108.97(3) |
| $As_2$-Fe-$As_3$, $\alpha$ (°) | 112.94(5) | 111.39(4) | 110.47(7) |
| $S_1$, Å | 2.377(4) | 2.302(4) | 2.273(4) |
| 1/2$S_2$/$h_P$, Å | 1.332(3) | 1.352(3) | 1.366(3) |
| $S_3$, Å | 1.714(4) | 1.688(4) | 1.668(4) |
| $S_4$, Å | 2.042(3) | 2.036(3) | 2.035(3) |
| Calculated density, g/cm$^3$ | 6.901 | 7.484 | 7.809 |
| Absorption coefficient, mm$^{-1}$ | 32.145 | 37.855 | 41.860 |
| F(000) | 863 | 892 | 906 |
| Crystal size, mm$^3$ | 0.11 x 0.07 x 0.03 | 0.11 x 0.06 x 0.03 | 0.11 x 0.08 x 0.03 |
| Theta range for data collection, deg. | 2.73 to 45.29 | 2.76 to 45.22 | 4.16 to 40.24 |
| Index ranges | -4=<h<=7, -7=<k<=8, -47=<l<=59 | -3=<h<=7, -6=<k<=7, -58=<l<=57 | -7=<h<=7, -7=<k<=4, -41=<l<=52 |
| Reflections collected/unique | 2778/670 $R_{int.}$= 0.0363 | 3827/640 $R_{int.}$= 0.0410 | 2986/490 $R_{int.}$= 0.0443 |
| Completeness to 2theta | 98.4 % | 97.3 % | 97.8 % |
| Data/restraints/parameters | 670/0/19 | 640/0/19 | 490/0/19 |
| Goodness-of-fit on $F^2$ | 1.436 | 1.224 | 1.187 |
| Final R indices [I>2sigma(I)] | $R_1$ = 0.0374, w$R_2$ = 0.0958 | $R_1$ = 0.0298, w$R_2$ = 0.0679 | $R_1$ = 0.0410, w$R_2$ = 0.0997 |
| R indices (all data) | $R_1$ = 0.0406, w$R_2$ = 0.0997 | $R_1$ = 0.0313, w$R_2$ = 0.0684 | $R_1$ = 0.0428, w$R_2$ = 0.1002 |

Fig. 1a

TABLE 2. Atomic coordinates and equivalent isotropic and anisotropic displacement parameters [Å$^2$ x 10$^3$] for the $L_4Fe_2As_2Te_{1-x}O_4$.

| Atom | Site | x | y | z L=Pr/Sm/Gd | $U_{iso}$ | $U_{11}=U_{22}$ | $U_{33}$ |
|---|---|---|---|---|---|---|---|
| $L_1$ | 4*e* | ½ | ½ | 0.0684(1)/0.0690(1)/0.0693(1) | 9(1)/7(1)/8(1) | 8(1)/6(1)/6(1) | 11(1)/8(1)/11(1) |
| $L_2$ | 4*e* | 0 | 0 | 0.1480(1)/0.1470(1)/0.1467(1) | 9(1)/7(1)/8(1) | 9(1)/6(1)/7(1) | 10(1)/8(1)/10(1) |
| As | 4*e* | ½ | ½ | 0.2054(1)/0.2042(1)/0.2035(1) | 12(1)/9(1)/12(1) | 13(1)/9(1)/13(1) | 10(1)/8(1)/9(1) |
| Te | 2*a* | 0 | 0 | 0 | 16(1)/13(1)/17(1) | 18(1)/15(1)/20(1) | 12(1)/10(1)/12(1) |
| Fe | 4*d* | ½ | 0 | ¼ | 11(1)/9(1)/10(1) | 10(1)/8(1)/8(1) | 12(1)/10(1)/13(1) |
| O | 8*g* | ½ | 0 | 0.1073(2)/0.1072(1)/0.1072(3) | 10(1)/8(1)/8(1) | 10(2)/6(1)/6(3) | 12(2)/12(1)/15(3) |

$U_{iso}$ is defined as one third of the trace of the orthogonalized $U_{ij}$ tensor. The anisotropic displacement factor exponent takes the form: $-2\pi^2$ [ ($h^2a^{*2}U_{11}$ + ... + $2hka^*b^*U_{12}$]. For symmetry reasons $U_{23}=U_{13}=U_{12}=0$.

TABLE 3. Details of the structure refinement for the $L_4Fe_2As_2Te_{1-x}O_{4-y}F_y$ (L = Sm, Gd) crystals.

| Empirical formula | $Sm_4Fe_2As_2Te_{0.72(1)}O_{4-y}F_y$ | $Gd_4Fe_2As_2Te_{0.92(1)}O_{4-y}F_y$ |
|---|---|---|
| $T_c$, K | 40 | 45 |
| Unit cell dimensions, Å, deg | $a$= 3.9597(5), $c$= 29.268(5) | $a$= 3.9363(6), $c$= 29.350(5) |
| Volume, Å$^3$ | 458.89(11) | 454.77(12) |
| $Ln_1$-$Ln_2$, Å | 3.6413(6) | 3.6124(10) |
| O-O= Fe-Fe, Å | 2.7999(4) | 2.7834(4) |
| $Ln_2$-$As_1$/$Ln_1$-Te, Å | 3.2424(8)/3.4406(5) | 3.2271(13)/3.4454(7) |
| $Ln_2$-O/$Ln_1$-O, Å | 2.319(3)/2.275(3) | 2.298(5)/2.263(5) |
| $As_1$-$As_2$, Å | 3.8956(12) | 3.9075(21) |
| Fe-As, Å | 2.3987(7) | 2.3987(12) |
| $As_1$-Fe-$As_2$, $\beta$ (°) | 108.59(2) | 109.07(4) |
| $As_2$-Fe-$As_3$, $\alpha$ (°) | 111.25(5) | 110.27(8) |
| $S_1$, Å | 2.330(4) | 2.301(4) |
| $S_2/h_{pn}$, Å | 1.355(3) | /1.370(3) |
| $S_3$, Å | 1.633(4) | 1.635(4) |
| $S_4$, Å | 1.999(4) | 2.031(4) |
| Calculated density, g/cm$^3$ | 7.373 | 7.854 |
| Absorption coefficient, mm$^{-1}$ | 37.614 | 41.938 |
| $F(000)$ | 871 | 910 |
| Crystal size, mm$^3$ | 0.024 x 0.118 x 0.179 | 0.033 x 0.054 x 0.065 |
| Theta range for data collection, deg. | 2.78 to 32.49 | 4.17 to 36.21 |
| Index ranges | -5=<h<=5, -5=<k<=5, -41=<l<=44 | -4=<h<=6, -6=<k<=6, -42=<l<=48 |
| Reflections collected/unique | 2293/301 $R_{int.}$= 0.0559 | 2294/390 $R_{int.}$= 0.0365 |
| Completeness to 2theta | 97.7 % | 97.5 % |
| Data/restraints/parameters | 301/0/20 | 390/0/19 |
| Goodness-of-fit on $F^2$ | 1.253 | 1.215 |
| Final R indices [I>2sigma(I)] | $R_1$ = 0.0251, w$R_2$ = 0.0557 | $R_1$ = 0.0422, w$R_2$ = 0.0982 |
| R indices (all data) | $R_1$ = 0.0252, w$R_2$ = 0.0558 | $R_1$ = 0.0433, w$R_2$ = 0.0987 |

(Fig. 1a)

TABLE 4. Atomic coordinates and equivalent isotropic and anisotropic displacement parameters [Å$^2$ x 10$^3$] for the $L_4Fe_2As_2Te_{1-x}O_{4-y}F_y$ (L = Sm, Gd) crystals.

| Atom | Site | x | y | Z L = Sm/Gd | $U_{iso}$ | $U_{11}=U_{22}$ | $U_{33}$ |
|---|---|---|---|---|---|---|---|
| $L_1$ | 4$e$ | ½ | ½ | 0.0683(1)/0.0692(1) | 10(1)/8(1) | 8(1)/6(1) | 13(1)/11(1) |
| $L_2$ | 4$e$ | 0 | 0 | 0.1479(1)/0.1476(1) | 10(1)/8(1) | 8(1)/8(1) | 12(1)/10(1) |
| As | 4$e$ | ½ | ½ | 0.2037(1)/0.2033(1) | 11(1)/11(1) | 11(1)/13(1) | 11(1)/8(1) |
| Te | 2$a$ | 0 | 0 | 0 | 16(1)/15(1) | 11(1)/17(1) | 11(1)/11(1) |
| Fe | 4$d$ | ½ | 0 | ¼ | 11(1)/10(1) | 14(1)/9(1) | 14(1)/12(1) |
| O | 8$g$ | ½ | 0 | 0.1066(2)/0.1072(2) | 9(1)/6(2) | 7(2)/3(4) | 16(1)/12(4) |